\documentclass[a4paper,11pt]{article}
\usepackage{pos}

\title{Parametrization sampling and the pion PDF in a phenomenological analysis}

\author*[a]{Aurore Courtoy}

\affiliation[a]{Instituto de F\'isica,
  Universidad Nacional Aut\'onoma de M\'exico, \\Apartado Postal 20-364,
  01000 Ciudad de M\'exico, Mexico}

\emailAdd{aurore@fisica.unam.mx}

\abstract{
In these proceedings, we extend the discussion of the pion PDF obtained by NLO QCD analysis in the Fant\^omas4QCD framework. Our pion analysis uses a state-of-the-art statistical methodology that accounts for the epistemic uncertainty. 
Fantômas4QCD, designed to handle multiple functional forms for solving the inverse problem, systematically explores a variety of solutions for PDFs,
thereby improving estimates of 
epistemic uncertainty.
Through this novel approach, we interpret our results for the valence sector, considering various non-perturbative methods available for predicting the pion PDF. We emphasize the distinctions between (global) QCD analyses and QFT-based calculations. 
}

\FullConference{10th International Conference on Quarks and Nuclear Physics (QNP2024)\\
8-12 July, 2024\\
Barcelona, Spain\\}

\begin{document}
\maketitle

\section{Goals of data-based QCD analyses}

Parton Distribution Functions (PDFs) are non-perturbative objects. They encapsulate the behavior of quarks and gluons inside hadrons for a given configuration of flavor, spin, and other relevant quantum numbers. PDFs follow a quantum-field theoretic definition, which is formally obtained through factorization theorems, at a given order in perturbation theory.
Due to their very nature, PDFs cannot be evaluated from first principles.
Since PDFs are of the utmost importance for predicting processes which involve hadron targets and/or beams, these functions have been determined through two main approaches, which we will call {\it phenomenological extractions} and {\it low-energy predictions}, nowadays complemented by lattice QCD approaches.

In the former approach, PDFs are extracted from the observables. Factorization theorems tell us how to implement the perturbative, calculable parts of a given observable.  Then, the extraction of the PDFs is performed by solving an inverse problem, sketched as follows for structure functions
\begin{equation}
\label{eq:DISfactorization}
F(x_{\rm B}, Q^2)=\sum_a\int_{x_{\rm B}}^1 \frac{dx}{x}f_{a/p}(x,\mu^2)\, {H_{a}}\left(\frac{x_{\rm B}}{x},\frac{\mu^2}{Q^2}\right)
+ {\cal O}\!\left({M}/{Q}\right)
\;,
\end{equation}
with $H_a$ the hard part of the observable, and $f_{a/p}(x)$ the PDF for a parton $a$ carrying a longitudinal momentum fraction $x$ of a hadron $p$.
DGLAP equations relate distribution functions at different resolution scales, $\mu^2$. Higher orders in perturbative QCD can be implemented in both factors, {\it i.e.} $H_a$ and $f_{a/p}$. Some higher-twist corrections are contained in the last term, ${\cal O}\!\left({M}/{Q}\right)$.
From the factorization point of view, all of the unknown, long-distance behavior is embedded into the PDF, up to power-suppressed corrections.
Since this methodology starts from observables to extract knowledge on a function by solving an inverse problem, we call it a {\sf top-down approach}.
The PDFs determined in the top-down approach are then used to make predictions for further observables. In the case of the proton PDFs, this is of high relevance for hadron colliders, such as the LHC, where PDFs are the baseline of most calculations.  More recently, global analyses of the pion PDFs have been resumed, thanks to theoretical and computational efforts of the past two decades.  
The pion PDFs shed light on the meson structure formation, the key aspect of understanding the low-energy regime of QCD. In these proceedings, we will discuss the {\sf top-down approach} to the pion PDF realized by the Fant\^omas4QCD project~\cite{Kotz:2023pbu},
complementing the panorama of data-based, top-down analyses of the pion PDFs by the JAM~\cite{Barry:2021osv} and the xFitter collaborations \cite{Novikov:2020snp}.

On the other hand, low-energy predictions provide an evaluation of the QFT definition of the PDFs, 
\begin{equation}
\label{eq:pionQFTPDF}
f_{a/p}(x,Q_0^2)=\int \frac{dz^-}{4\pi}   e^{-i (x-\frac{1}{2}) P^+z^-}
\left\langle \pi^+(p)\right|\bar{\psi}_a\left(0,\frac{y^-}{2}, {\bf 0}\right) \gamma^+ \psi_a\left(0,-\frac{y^-}{2}, {\bf 0}\right)\left|\pi^+(p)\right\rangle
\;,
\end{equation}
here for the $\pi^+$ and in the light-cone gauge ($A^+=0$),
incorporating (state-of-the-art) calculations of dressed propagators and vertex functions that reflect both the pseudo-Goldstone and bound-state nature of the pion, within accessible hypotheses, {\it e.g.},~\cite{Hecht:2000xa,Davidson:2001cc,Theussl:2002xp,Ding:2019lwe,Lan:2024ais}. 
The variable $Q_0^2$ is a low, hadronic scale at pre-factorization values, where the degrees of freedom of the non-perturbative calculation are expected to connect to the RGE spectrum in the $\overline{MS}$ scheme. 
We call this the {\sf bottom-up approach}.

The results from the  bottom-up approach are frequently compared with those from the  top-down approach. This raises the question of whether such a comparison is meaningful. This issue will be revisited shortly.

\section{Fant\^omas in a nutshell}

In the  top-down narrative, we, the practitioners, face an inverse problem that involves the determination of a function of $x$ from a finite number of discrete data points, through a convolution. This problem admits more than one solution for the PDFs achieving a good agreement with the data. In order to investigate the impact of multiple solutions, the Fant\^omas4QCD project was created, based on previous studies of functional mimicry~\cite{Courtoy:2020fex}  and representative sampling~\cite{Courtoy:2022ocu} in the context of PDFs. We parametrize the PDFs as the product of a carrier function, which describes the asymptotics, and a modulator, which we take to include a B\'ezier curve of degree $N_m$, ${\cal B}^{(N_m)}$,
\begin{eqnarray}
    x\,f_i(x, Q_0^2)= A_i x^{B_i} (1-x)^{C_i} \left[ 1+ {\cal B}^{(N_m)}(y(x)) \right].
    \label{eq:xfpion} 
\end{eqnarray}
The novelty in our {\sf metamorph} function, Eq.~(\ref{eq:xfpion}), is that the B\'ezier curve is directly related to its values $P_j$ at control points $x_j$, 
i.e. $P_j =  {\cal B}^{(N_m)}(y(x_j))$, with $y(x)$ a function of $x$. If we use $N_m+1$ such control points, the vector of coefficients of the polynomial can uniquely be determined by a simple matrix equation, see Ref.~\cite{Kotz:2023pbu} and references therein. The {\sf metamorph} captures the modulations in the PDF due to the shift of the control points. By varying the position and the number of fixed control points or the function $y(x)$, we are able to generate, on the fly,  many different functional forms for the {\it l.h.s.} of Eq.~(\ref{eq:xfpion}).

\begin{figure}[bt]
\begin{center}
    \includegraphics[width=0.475\columnwidth]{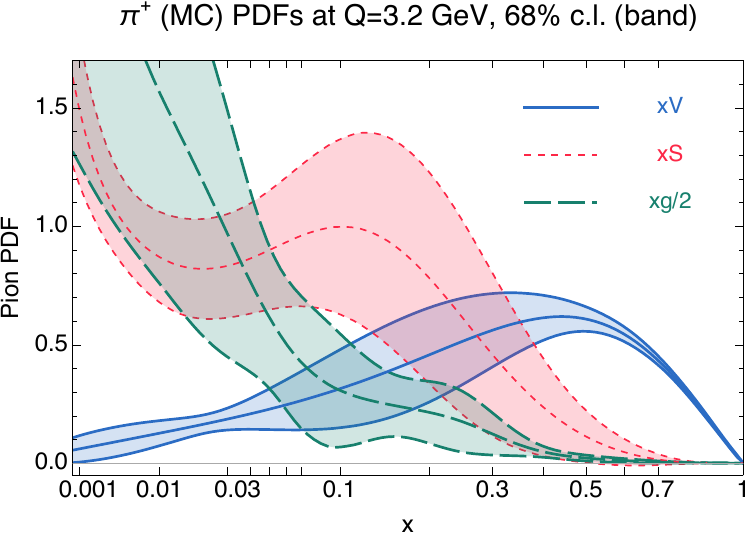}
    \includegraphics[width=0.475\columnwidth]{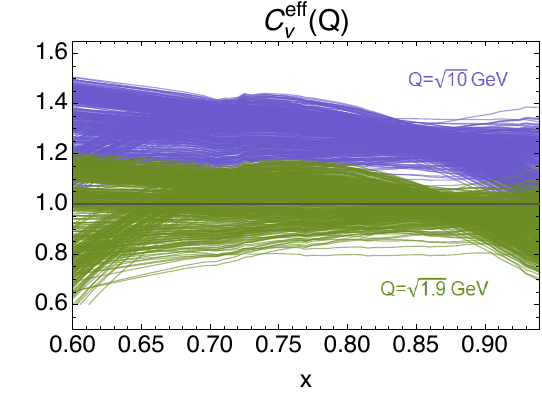}
    \caption{ Left: Fant\^oPDFs for the valence (blue, full curves), the sea (red, dotted curves) and gluon (green, dashed curves) at $Q=\sqrt{10}$ GeV.
    Right: The effective $(1-x)$ exponent of the valence PDF in the Fant\^oPDF
      ensemble at $Q_0=\sqrt{1.9}$ GeV (green), and   at
      $\sqrt{10}$ GeV (blue). 
      } 
    \label{fig:a2eff}
\end{center}
\end{figure}

This innovative parametrization was used to fit the pion PDF using the xFitter framework. xFitter  already contained the pion-induced Drell-Yan data as well as prompt photon's, to which we have added a minimal set of leading-neutron data. The Drell-Yan data largely constrains the large-$x$ valence and sea quark pion PDF, while the other two bring more information on the sea and the gluon, though not enough presently to disentangle them. 
The Fant\^omas framework  allows, for the first time, for the accounting of the statistical ({\it aleatoric}) uncertainties coming from the data but also the systematics that now include methodological choices and the quality of various samplings, known as {\it epistemic} uncertainties. That is, sampling over the space of solutions for the function of interest contributes to the {\it epistemic uncertainty}. On the {\it l.h.s.} of Fig.~\ref{fig:a2eff}, we show the resulting combination of the 5 most diverse solutions, after $\sim 100$ trials filtered through soft constraints, for the valence, sea and gluon PDFs of the pion.

\section{The valence sector of the pion PDF-- a critical view}

Having tried approximately a hundred different functional forms, we can conclude on the size of the uncertainties in given regions of the $(x, Q^2)$ plane. The very large-$x$ and moderate $Q^2$ region can be compared against early-QCD predictions. Counting rules suggested a  large-$x$ behavior of the pion quark PDF of $\lim_{x\to 1}f_V^{\pi}(x)\propto(1-x)^\beta$ with $\beta \approx 2$, in the scaling region.
 This expectation  
 may be modified by various radiative contributions at 
large momentum fractions that affect the interpretation of realistic measurements~\cite{Courtoy:2020fex}. 
In this regard, the  Fant\^omas4QCD analysis did not qualitatively differ from  other recent analyses: the fall-off of the valence PDF at large $x$ is compatible with $\beta=C_V^{\mbox{\tiny eff}}=1$ at
 $Q_0=\sqrt{1.9}$ GeV, in spite of the
 multiple functional forms that have been considered (Fig.~\ref{fig:a2eff} right). 
 Threshold effects could affect the extraction of the pion PDF at large~$x$. The JAM collaboration investigated various treatments of resummation and found that $\beta$ could  significantly vary, and  chose to adopt the result that leads to $\beta \approx 1$, motivated by the most accepted resummation treatment~\cite{Barry:2021osv}. Hence comes the question of comparison of predictions with data-driven analyses.

In Ref.~\cite{Courtoy:2020fex}, we argued that polynomial mimicry implies that there is no sufficient condition to confirm that one specific non-perturbative representation of a PDF is the true one when compared to a phenomenologically extracted polynomial, or even compared against data.
A necessary condition is that the curves agree according to an agreed-upon metric,
such as $\chi^2$. 
For practical use of the  bottom-up results, the hadronic scale $Q_0^2$ must be determined. It is customary to fix it by comparison of the {\it l.h.s.} of Eq.~(\ref{eq:pionQFTPDF}) with observables (see, {\it e.g.} Ref.~\cite{Ceccopieri:2018nop}) or moments obtained in global analyses (see, {\it e.g.} Ref.~\cite{Stratmann:1993aw}), by using the backward DGLAP evolution. Since $Q_0^2$ turns out to be of a few hundred MeV, this procedure relies on pushing the validity of DGLAP evolution to extremely low scales. Once the hadronic scale is determined, it is also used to make predictions on functions or observables, using forward DGLAP evolution. Predictions made using this method do not have a straightforward interpretation. 

\begin{figure}[t]
\begin{center}
    \includegraphics[width=0.7\columnwidth]{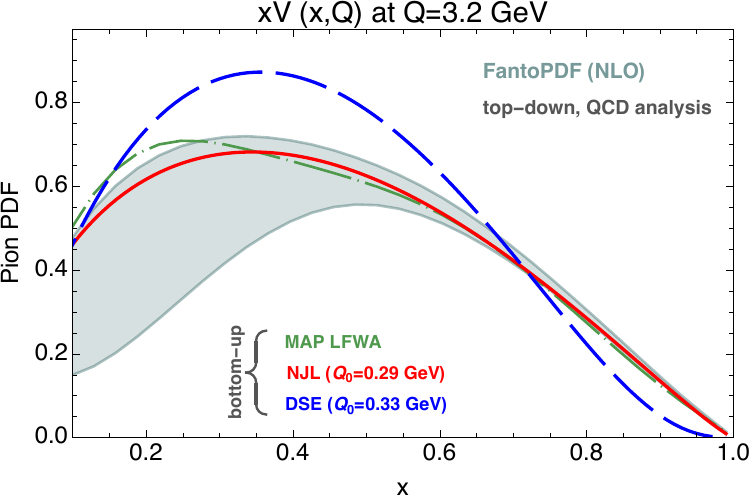}
    \caption{The valence  of the pion at $Q=\sqrt{10}$ GeV, in linear scale.
    The central value of MAP LFWA result is shown  in green (long-dashed–dotted curve), the NJL result evolved for ($Q_0 = 0.29, \Lambda_{\rm LO} =
0.2$) GeV, in red (full curve), and the DSE result of Ref.~\cite{Ding:2019lwe}  for ($Q_0 = 0.33, \Lambda_{\rm LO} =
0.2$) GeV, in blue (dashed curve), and compared to the Fant\^omas PDFs, shown in dark cyan (solid band).}
    \label{fig:fantomod}
\end{center}
\end{figure}

For illustration purposes, we present in Fig.~\ref{fig:fantomod} the central values of na\" ively evolved  bottom-up results alongside the comprehensive global  top-down QCD analysis.
In this figure, both 
 the Nambu--Jona-Lasinio model~\cite{Davidson:2001cc,Theussl:2002xp}  (labeled ``NJL") and the Dyson-Schwinger inspired analysis of Ref.~\cite{Ding:2019lwe} (labeled ``DSE") are evolved from a very low hadronic scale ($Q_0=0.29$ and $0.33$ GeV, respectively) to $Q=\sqrt{10}$ GeV using the LO DGLAP evolution as implemented in Refs.~\cite{Block:2009en,Golec-Biernat:1998zbo}. Thus, in Fig.~\ref{fig:fantomod}, $xV(x, Q_0)_{\rm bottom-up}$ serves solely as an initial condition for the LO DGLAP equations.
 Additionally, we show the central value
of the hybrid analysis using Light-Front Wave Functions~\cite{Pasquini:2023aaf} (labeled ``MAP LFWA"), which is a NLO analysis providing its own LHAPDF grids.

The interpretation of Fig.~\ref{fig:fantomod} is as follows. The MAP LFWA free parameters are determined  from fitting the Drell-Yan and prompt-photon data of xFitter's NLO framework. The mid- and large-$x$ behavior of the central value of MAP valence PDF follows that of the Fant\^oPDFs's.
As anticipated, the predictions
of NJL and DSE cannot be directly  compared to the data-driven, fixed-order $\overline{MS}$ PDFs of the Fant\^omas analysis. The uncertainty on the formers is unknown, due to the non-perturbative nature of the approaches. What the red and blue curves reflect is the evolution of a flat and a quadratic PDF as initial conditions, respectively,  with LO DGLAP equation. Variations of the value of $Q_0$, in our working hypotheses, do not allow for a transition from one initial condition to the other.

\section{Conclusions}

In this contribution to the proceedings, we have discussed the recent results on the PDFs of the pion with epistemic uncertainties. A representative sampling of the parametrization, a solution to the inverse problem of interest, increases the size of the uncertainties of the valence, sea and gluon PDFs in most of the $x$ range. However, since the Drell-Yan data constrains the large-$x$ region greatly, our fixed-order analysis agrees with previous extractions in the limit $x\to 1$. 

We commented on the difference between global QCD { top-down} analyses and non-perturbative { bottom-up} approaches.
For phenomenological PDFs to meet the predictions from non-perturbative methodologies, further studies must be undertaken. 

\section*{Acknowledgments}

AC would like to thank P. Nadolsky for fruitful discussions, S. Noguera for providing the code for the DGLAP evolution of Ref.~\cite{Block:2009en}, as well as useful discussions on the NJL results,  J. Segovia for useful exchanges, and S. Venturini for providing the MAP LFWA~\cite{Pasquini:2023aaf} grids. 
AC is supported by the UNAM Grant DGAPA-PAPIIT IN111222 and
CONACyT Ciencia de Frontera 2019 No. 51244 (FORDECYT-PRONACES).

\end{document}